\documentclass[preprint]{aastex}

\slugcomment{}

\shorttitle{BLACK HOLE MAGNETOSPHERES}
\shortauthors{TOMIMATSU \& TAKAHASHI}

\begin{document}

\title{BLACK HOLE MAGNETOSPHERES AROUND THIN DISKS \\
        DRIVING INWARD AND OUTWARD WINDS}

\author{AKIRA TOMIMATSU\altaffilmark{}}
\affil{Department of Physics, Nagoya University, Chikusa-ku, Nagoya 
464-8602, Japan}
\email{atomi@allegro.phys.nagoya-u.ac.jp}

\and

\author{MASAAKI TAKAHASHI\altaffilmark{}}
\affil{Department of Physics and Astronomy, Aichi University of Education, 
Kariya, Aichi 448-8542, Japan}
\email{takahasi@phyas.aichi-edu.ac.jp}

\begin{abstract}
We construct a simple model for stationary, axisymmetric black-hole 
magnetospheres, in which the poloidal magnetic field is generated by a 
toroidal electric current in a thin disk with the inner edge, by solving 
the vacuum Maxwell equations in Schwarzschild background. In this work, to 
obtain a concise analytical form of the magnetic stream function, we use 
the approximation that the inner edge is far distant from the event 
horizon. The global magnetospheric structure with the closed-loop and open 
field lines threading the inner and outer parts of the disk is explicitly 
shown, claiming that the model is useful as a starting point to study 
astrophysical problems involving inward disk-driven winds to a black hole 
and outward ones to infinity. The asymptotic shape of the field lines at 
the event horizon becomes nearly cylindrical, while at infinity it becomes 
conical. The magnetic spot in the disk connected with the black hole 
through the loop field lines occupies a very narrow region with the ring 
area roughly equal to the horizon area. By taking account of the existence 
of a uniform (external) magnetic field, we also obtain the model for 
collimated open field lines. Then, it is found that the magnetic connection 
between the black hole and the disk breaks down if the uniform field is 
strong enough. Considering slow rotation of the magnetosphere and angular 
momentum transfer by inward winds from the disk, the final discussion is 
devoted to gradual disruption of the closed loops due to radial accretion 
of disk plasma toward the black hole.
\end{abstract}


\section{INTRODUCTION}

It is widely believed that a supermassive black hole surrounded by an 
accretion disk can work as a central engine of active galactic nuclei. If 
magnetic field exists around the black hole and the disk, it is important 
to understand the magnetospheric structure for explaining the highly 
energetic phenomena. The magnetohydrodynamical (MHD) approach in general 
relativity may be necessary to study various features of the black hole 
magnetospheres. Unfortunately, even if the magnetosphere is assumed to be 
stationary and axisymmetric, any global models describing both inward winds 
in the strong-gravity region near black hole and outward winds in the far 
distant region have not been constructed, in spite of a large number of 
works devoted to the MHD problems. This is mainly because it is very 
difficult to analyze the highly nonlinear Grad-Shafranov equation for the 
magnetic stream function which defines the poloidal field lines (see e.g., 
\citet{bes97} for review).

To avoid the mathematical difficulty in the full MHD system, one may 
consider vacuum solutions of the magnetic stream function as a practical 
approximation applicable to the magnetically dominated models. In 
particular, the well-known examples such as split-monopole, uniform and 
paraboloidal fields in Schwarzschild geometry \citep{mi73,w74,bz77} have 
been used as the zeroth-order approximation for the perturbation method in 
slow-rotation and force-free fields, from which the first-order 
perturbation has been derived for giving a weak toroidal magnetic field and 
discussing the Blandford-Znajek process \citep{bz77}. Another important MHD 
problem to be studied is fluid motion along poloidal field lines. For 
examples, the existence of accretion flows onto the black hole passing 
through some MHD critical points has been shown under a fixed shape of 
field lines \citep{tak90}. The vacuum solutions allowing a field 
configuration of astrophysical interest will become a useful tool in an 
attempt to discuss the trans-critical motion of outgoing and ingoing 
magnetized winds, except in the region where the effect of fluid inertia 
should dominate to change crucially the field structure.

Of course, even for the vacuum fields, some current distribution should be 
assumed to exist in a restricted region of the magnetosphere. In this paper 
we focus on the magnetic field generated by a toroidal current distribution 
in a thin disk around the Schwarzschild black hole. The source of the 
above-mentioned fields with split-monopole and paraboloidal structures can 
be also disk currents. However, such models contain only open field lines 
and fail to describe a magnetic connection of the disk with the black hole. 
This will be partially due to the absence of the disk inner edge separated 
from the event horizon. The magnetic connection by closed field lines is 
expected to exist and have astrophysically important effects 
\citep{pu91a,gru99,bl00,li00c,kl00}. Hence, our purpose here is to present 
explicitly a global magnetospheric model with both closed and open field 
lines threading a thin disk and connecting magnetically the inner and outer 
parts, respectively, with the black hole and the far distant region, which 
is relevant to the problems of inward and outward disk-driven winds. Though 
the vacuum model would miss some important features of more realistic 
magnetospheres, our modeling could be a preliminary step for understanding 
the various MHD processes in further detail.

To obtain general models based on a disk current distribution, it may be 
convenient to consider linear superposition of the magnetic field generated 
by a single ring current located at a fixed radius on the equatorial plane 
which has been found in Kerr geometry \citep{pet75}. Assuming the current 
distribution from an inner edge to an outer one, \citet{li00a} has derived 
a formal expression of the magnetic stream function in terms of the 
infinite sum of the multipole fields, without revealing the global shape of 
the field lines. Because the each multipole field of which the structure 
has been discussed by \citet{gos00} still fails to give closed field lines 
regular at the event horizon, the infinite sum is essential to 
understanding the global shape.

In \S2 we would like to develop a different approach to arrive at the 
infinite sum formula for the magnetic stream function. In \S3 we succeed in 
reducing it to an analytical simple form by locating the disk inner edge at 
a radius much larger than the horizon radius and the outer edge at infinite 
radius. The global structure of closed and open field lines is shown, and 
it is found that only the very narrow region of the disk separated from the 
inner edge can be magnetically connected with the black hole, together with 
the result that the field lines become nearly cylindrical at the event 
horizon and conical at infinity. In \S4 we also consider a modification of 
the global shape due to an addition of a uniform external field. Though 
collimated field lines appear at the far distant region, it is found that 
for a strong uniform field the magnetic connection between the black hole 
and the disk breaks down. Finally we discuss the MHD effect due to slow 
rotation of the magnetosphere (namely, angular momentum transfer by 
magnetically dominated inward winds from the disk to the black hole), which 
could lead to radial accretion of disk plasma and gradual disruption of the 
closed field lines \citep{pc89}. In the following we use the geometrical 
units with $G=c=1$.

\section{THE MAGNETIC STREAM FUNCTION}

Let us consider a stationary, axisymmetric magnetic field in Schwarzschild 
background of the metric
\begin{equation}
  ds^{2} \ = \ -\left(1-\frac{2M}{r}\right) dt^{2}
               +\left(1-\frac{2M}{r}\right)^{-1}dr^{2} 
               +r^{2}\left(d\theta^{2}+\sin^{2}\theta d\phi^{2}\right) \ ,
\end{equation}
where $M$ is the mass of the black hole. The poloidal components of the 
magnetic field may be given by
\begin{equation}
     \vec{B}_{p} \ = \ 
     \frac{\vec{\nabla}\Psi\times\vec{\phi}}{2\pi r\sin\theta} \ ,
\end{equation}
where $\vec{\phi}$ is a unit vector in the azimuthal direction, and 
$\Psi(r,\theta)$ is the so-called magnetic stream function defining the 
poloidal field lines by $\Psi=$ \ constant. If the vacuum field is assumed, 
the Maxwell equations reduce to the simple form
\begin{equation}
  r^{2}\frac{\partial}{\partial r}\left[\left(1-\frac{2M}{r}\right)
  \frac{\partial\Psi} {\partial r}\right] 
  +\sin\theta\frac{\partial}{\partial\theta} \left(\frac{1}{\sin\theta}
  \frac{\partial\Psi}{\partial\theta}\right) \ = \ 0 \ ,
\end{equation}
which allows the separable solutions
\begin{equation}
  \Psi \ = \ R_{\nu}(r)\Theta_{\nu}(\theta) \ .
\end{equation}
with the separation constant $\nu$

For the analysis of these functions it is convenient to introduce the variables
\begin{equation}
  w \ = \ \frac{r}{M}-1 \ , \ \ x \ = \ \cos\theta \ .
\end{equation}
Then, from the regularity at the polar axis $\theta=0, \pi$  we obtain the 
angular part written by
\begin{equation}
  \Theta_{\nu} \ = \ \int_{|x|}^{1} P_{\nu}(x')dx' \ ,
\end{equation}
where $P_{\nu}$ is the Legendre function of the first kind. Though in this 
paper we focus on the field with equatorial symmetry, the parameter $\nu$ 
is not limited to a positive odd integer for allowing a discontinuous 
change of the radial component of $\vec{B}_{p}$ on the upper and lower 
surfaces of the equatorial disk.

To determine the radial part $R_{\nu}$, the condition for regularity at the 
event horizon $r=2M$ and at infinity $r\rightarrow\infty$ becomes 
important. We use the Legendre function $P_{\nu}(w)$ to give the solution 
regular at $w=1$ of the form
\begin{equation}
  R_{\nu} \ = \ (w+1)P_{\nu}(w)-\int_{1}^{w}P_{\nu}(w')dw' \ ,
\end{equation}
where we assume to be $\nu=(-1/2)+ik$ with arbitrary positive $k$, because 
the oscillatory real function $P_{\nu}(w)$ has the amplitude decreasing in 
proportion to $1/\sqrt{w}$ at large $w$, and the linear superposition
\begin{equation}
  \Psi(r,\theta) \ = \ \int_{0}^{\infty} f(k)R_{\nu}(r)\Theta_{\nu}(\theta) dk
\end{equation}
will be able to describe a global magnetic field satisfying the boundary 
conditions.

Now we calculate the coefficient $f(k)$ in the linear superposition 
corresponding to the case of a disk current with the inner edge at 
$w=w_{0}>1$. Recall that in previous works the solution was written by an 
infinite sum of the multipole fields with $\nu$ of natural numbers. Then, 
in the distant region $w>w_{0}$ the Legendre function $P_{\nu}(w)$ should 
be replaced by $Q_{\nu}(w)$ with a change of the coefficient due to the 
condition for continuity of $\Psi(r,\theta)$ at $w=w_{0}$. We can avoid 
these slightly complicated steps by using the integral form (8). Though the 
toroidal surface current in the thin disk is proportional to 
$r^{-2}(\partial\Psi/\partial\theta)$ at $\theta=\pi/2$, here we rather 
focus on the equality
\begin{equation}
  \frac{\partial}{\partial r}\left(\frac{\partial\Psi}
    {\partial\theta}\right) \ = \ \frac{w+1}{M}\frac{d}{dw}h(w) \ ,
\end{equation}
which is evaluated at the upper surface of the equatorial plane. The 
remarkable point is that the function $h(w)$ is given by
\begin{equation}
  h(w) \ = \ \int_{0}^{\infty}g(k)P_{\nu}(w) dk \ ,
\end{equation}
where we have
\begin{equation}
  g(k) \ = \ P_{\nu}(0)f(k) \ ,
\end{equation}
and the inversion formula \citep{erd53} holds as follows,
\begin{equation}
  g(k) \ = \ k\tanh(\pi k)\int_{1}^{\infty}P_{\nu}(w)h(w) dw \ .
\end{equation}
Then, if the function $h(w)$ is specified according to the current 
distribution, we can derive the coefficient $f(k)$ giving the magnetic 
stream function $\Psi(r,\theta)$ valid in the whole region.

Further, to obtain the explicit form of $f(k)$ through the well-known 
integral formula involving two Legendre functions, let us assume in the 
disk region $w>w_{0}$ to be
\begin{equation}
  h(w) \ = \ A_{0}Q_{s}(w) \ .
\end{equation}
Though $A_{0}$ may be an arbitrary real parameter, we limit $s$ to be 
positive for assuring that $\partial\Psi/\partial\theta$ does not diverge 
at large $w$, where the Legendre function $Q_{s}(w)$ of the second kind 
falls off as $w^{-s-1}$. (If any different model of the current 
distribution becomes necessary, the linear superposition of $Q_{s}$ with 
various $s$ in the range $0<s<\infty$ is a possible procedure to construct 
the solution $\Psi$. However, we would like to emphasize that this simple 
example for the current distribution can be useful for revealing 
interesting features of the magnetospheric structure, as will be shown in 
the following.) Because in the inner region $1\leq w<w_{0}$, where the 
field lines should vertically thread the equatorial plane, we have
\begin{equation}
  h(w) \ = \ 0 \ ,
\end{equation}
it is easy to arrive at the result
\begin{equation}
  g(k) \ = \ -\frac{A_{0}k\tanh(\pi k)}{k^{2}+[s+(1/2)]^{2}}F_{\nu} \ ,
\end{equation}
where
\begin{equation}
  F_{\nu} \ = \   (w_{0}^{2}-1)\left[P_{\nu}(w_{0})
  \frac{dQ_{s}(w_{0})}{dw_{0}}-Q_{s}(w_{0})
  \frac{d P_{\nu}(w_{0})}{dw_{0}}\right] \ .
\end{equation}
Strictly speaking, even if the condition (14) is assumed, equation (9) only 
implies that $\partial\Psi/\partial\theta$ is constant at the upper or 
lower surface of the equatorial plane in the inner range $1\leq w<w_{0}$. 
Hence, we must confirm the validity of equation (15), by calculating the 
value on the event horizon $w=1$, which is given by
\begin{equation}
  \frac{\partial\Psi}{\partial\theta} \ = \ \int_{0}^{\infty}2g(k)dk \ ,
\end{equation}
at the upper surface of the equatorial plane. In fact, the right-hand side 
of equation (17) is shown to vanish by virtue of the equality
\begin{equation}
  Q_{s}(w_{0}) \ = \ \int_{0}^{\infty}\frac{k\tanh(\pi k)}
                   {k^{2}+[s+(1/2)]^{2}}P_{\nu}(w_{0})dk
\end{equation}
which is a result of the inversion formula (12).

We have obtained the magnetic stream function of the integral form (8) in 
which the coefficient $f(k)=g(k)/P_{\nu}(0)$ is given by equation (15). To 
calculate the integral with respect to $k$ with the application of the 
residue theorem, it will be useful to consider the integration contour 
along the real axis ($-\infty<k<\infty$) and the large semi-circle in the 
complex $k$-plane, based on the relation between two Legendre functions 
written by
\begin{equation}
  P_{\nu}(z) \ = \ \frac{i}{\pi}\coth(\pi k)
  \left[Q_{\nu}(z)-Q_{-\nu-1}(z)\right] \ .
\end{equation}
If $w<w_{0}$ we replace $P_{\nu}(w_{0})$ in the integrand according to 
equation (19), and the large semi-circle is chosen to be in the lower (or 
upper) half of the complex k-plane for the integral of the term involving 
$Q_{\nu}$ (or $Q_{-\nu-1}$). Then, by taking account of the contribution 
only from the poles present at $k=\pm i[s+(1/2)]$ and at $k=\pm 
i[2n+(3/2)]$ where $P_{\nu}(0)$ vanishes for $n=0,1,2,\cdots$, we obtain 
the result valid in the inner region $1\leq w<w_{0}$ as follows,
\begin{equation}
  \Psi \ = \ 
  \sum_{n=0}^{\infty}\alpha_{n}G_{2n+1}R_{2n+1}(r)\Theta_{2n+1}(\theta) \ ,
\end{equation}
where
\begin{equation}
  \alpha_{n} \ = \ 
  \frac{A_{0}(-1)^{n}(4n+3)}{(2n+s+2)(2n+1-s)}\frac{(2n+1)!!}{(2n)!!}
\end{equation}
and
\begin{equation}
  G_{2n+1} \ = \ (w_{0}^{2}-1) 
  \left[ Q_{2n+1}(w_{0})\frac{dQ_{s}(w_{0})}{dw_{0}}-Q_{s}(w_{0})
  \frac{dQ_{2n+1}(w_{0})}{dw_{0}} \right] \ .
\end{equation}

In the outer region $w>w_{0}$, however, we must use equation (19) for 
$P_{\nu}(w)$ [instead of $P_{\nu}(w_{0})$] with the same choice of the 
large semi-circle, and the contribution from the poles is divided into the 
two parts
\begin{equation}
  \Psi \ = \ \Psi_{1}+\Psi_{2}
\end{equation}
with the forms written by
\begin{equation}
  \Psi_{1} \ = \ 
  \sum_{n=0}^{\infty}\alpha_{n}\Theta_{2n+1}(\theta)
  \left[ F_{2n+1}\tilde{R}_{2n+1}(r)
        -G_{2n+1}\int_{1}^{w_{0}}P_{2n+1}(w')dw' \right] \ ,
\end{equation}
and
\begin{equation}
  \Psi_{2} \ = \ 
  \frac{A_{0}}{\sqrt{\pi}}\Gamma\left(\frac{1-s}{2}\right)
  \Gamma\Biggl(1+\frac{s}{2}\Biggr)\Theta_{s}(\theta)\tilde{R}_{s}(r)
\end{equation}
where the function $\tilde{R}_{\nu}(r)$ for any $\nu$ is defined by
\begin{equation}
  \tilde{R}_{\nu}(r) \ = \ (w+1)Q_{\nu}(w)-\int_{w_{0}}^{w}Q_{\nu}(w')dw' \ ,
\end{equation}
and $\Gamma(z)$ is the gamma function.

It is easy to check that the continuity of these expressions (20) and (23) 
for $\Psi$ at the boundary $w=w_{0}$ holds, because through the application 
of the expansion in terms of the Legendre polynomials to the function 
$u(x)$ defined by $u=P_{s}(x)$ for $x>0$ and $u=-P_{s}(|x|)$ for $x<0$ we 
obtain
\begin{equation}
  \sum_{n=0}^{\infty}\frac{(-1)^{n}(4n+3)}{(2n+s+2)(2n+1-s)}\frac{(2n+1)!!}
  {(2 n)!!}P_{2n+1}(x) \ = \ 
  \frac{1}{\sqrt{\pi}}\Gamma\left(\frac{1-s}{2}\right)
  \Gamma\Biggl(1+\frac{s}{2}\Biggr)u(x) \ .
\end{equation}
Further, equation (27) is useful to show that for a positive $s$ the 
asymptotic behavior of $\Psi$ at $w\rightarrow\infty$ becomes
\begin{equation}
  \Psi \ \simeq \ -A_{0}\int_{w_{0}}^{\infty}Q_{s}(w')dw' \times v(x) \ ,
\end{equation}
where the dependence on $x$ is given by
\begin{equation}
  v(x) \ = \ 
  \sum_{n=0}^{\infty}\frac{(-1)^{n}(4n+3)}{(2n+1)(2n+2)}
  \frac{(2n+1)!!}{(2n)!!}\int_{|x|}^{1}P_{2n+1}(x')dx' \ .
\end{equation}
By substituting the value of $s=0$ into the equality (27), we can calculate 
the summation to be $v=1-|x|$. Hence, we conclude that this black hole 
magnetosphere based on a disk current model has the asymptotically conical 
shape of magnetic field lines. This asymptotic structure is a consequence 
of the disk currents extending to infinite distance. Though such a current 
distribution is unphysical, the existence of open field lines will be 
useful to discuss the problem of outgoing winds from the disk to infinity 
in our modeling.

\section{THE GLOBAL STRUCTURE OF FIELD LINES}

Though the asymptotic behavior of $\Psi$ at infinity is clear, the 
expression written by the infinite sum of the multipole fields is still 
complicated for understanding the features of the field line structure near 
the inner edge and the event horizon. Hence, let us assume that the inner 
edge is far distant from the event horizon. For example, the inner edge of 
the disk around the Schwarzschild black hole may be located at the 
innermost stable circular orbit $r=6M$ on the equatorial plane. Then, we 
have $w_{0}=5$, for which the approximation $w_{0}\gg 1$ will be roughly 
allowed.

In the following we focus on equation (20), because the multipole sum 
should arrive at the same result even if equation (23) is used. By virtue 
of the approximation $w_{0}\gg 1$, we obtain the dominant behavior of the 
Legendre function such that $Q_{s}(w_{0})\sim w_{0}^{-s-1}$. Except in the 
region near the event horizon where $w$ is of order unity, we can calculate 
equation (20) under the assumption $1\ll w\leq w_{0}$, which leads to the 
approximate relation
\begin{equation}
  P_{2n+1}(w)Q_{2n+1}(w_{0}) \ \simeq \ 
  \frac{2n+1}{(2n+2)(4n+3)}\left(\frac{w}{w_{0}}\right)^{2n+2} \ .
\end{equation}
Of course, for the angular part involving the integral of $P_{2n+1}(x)$ we 
cannot use such an approximation. Hence, our key step is to rewrite the 
Legendre polynomial through the integral formula
\begin{equation}
  P_{2n+1}(\cos\theta) \ = \ 
  \frac{1}{\pi}\int_{0}^{\pi}(\cos\theta+i\sin\theta\cos\varphi)^{2n+1}
  d\varphi \ ,
\end{equation}
which allows us to obtain
\begin{equation}
  \Psi \ \simeq \ A_{0}Q_{s}(w_{0})w\int_{|x|}^{1}dx' \ 
  \left[\frac{1}{\pi}\int_{0}^{\pi}d\varphi K(y) \right] \ .
\end{equation}
The variable $y$ and the function $K$ are defined by
\begin{equation}
  y \ = \ \frac{w}{w_{0}}\left[x'+i\cos\varphi\sqrt{1-(x')^{2}}\right] \ ,
\end{equation}
and
\begin{equation}
  K \ = \ 
  \sum_{n=0}^{\infty}\frac{(-1)^{n}(2n+1)}{2n+2+s}\frac{(2n+1)!!}{(2n+2)!!}y^{ 
2n+1} \ ,
\end{equation}
for which the infinite sum becomes possible to give
\begin{equation}
  K \ = \ 
  \frac{1}{y}\left(1-\frac{1}{\sqrt{1+y^{2}}}\right) 
  -(s+1)y^{-s-1}\int_{0}^{y}(y')^{s-1} 
  \left[ 1-\frac{1}{\sqrt{1+(y')^{2}}} \right] dy' \ .
\end{equation}
This expression of $K(y)$ is also valid in the outer region $w>w_{0}$, and 
it is easy to verify that at infinity where $y\gg 1$ the magnetic stream 
function given by equation (32) can keep the same asymptotic form as 
equation (28).

Next, let us give the magnetic stream function near the event horizon. 
Because $w$ is of order unity, the term with $n=0$ dominates in the 
multipole sum (20) under the approximation $w_{0}\gg 1$, and we have
\begin{equation}
  \Psi \ \simeq \ \frac{A_{0}Q_{s}(w_{0})}{4(s+2)w_{0}}(1-x^{2})(w+1)^{2} \ ,
\end{equation}
which shows the cylindrical magnetic field written by $\Psi=\pi 
B_{0}r^{2}\sin^{2}\theta$ with the strength
\begin{equation}
  B_{0} \ = \ \frac{A_{0}Q_{s}(w_{0})}{4\pi M^{2}(s+2)w_{0}} \ .
\end{equation}
This is in accordance with the result claimed for general multipole 
expansions by \citet{ki75}. However, if $w_{0}$ is of order unity, the 
dominance of the cylindrical field in the sum (20) apparently breaks down 
even near the event horizon, and a more complicated structure of field 
lines will become possible there.

We hereafter fix $B_{0}$ to be positive in equation (37). Then, for $s>0$, 
the magnetic stream function becomes negative at infinity, where we have
\begin{equation}
  \Psi \ \simeq \ -\left(1+\frac{2}{s}\right) 4\pi B_{0}(Mw_{0})^{2}
  \left(1-|x|\right) \ .
\end{equation}
Because this change of $\Psi$ occurs for all $\theta$, a closed field line 
of $\Psi=0$ connecting the equatorial plane with the polar axis should 
exist in the magnetosphere. This critical field line can thread the black 
hole along the polar axis $|x|=1$ and divide the magnetosphere into the 
inner region with closed field lines and the outer region with open field 
lines. Note that if equation (32) is written by the sum of the multipole 
fields we obtain the $r$-dependence at $x=0$ as follows,
\begin{equation}
  \Psi \ \simeq \ 
  B_{0}(Mw)^{2}\sum_{n=0}^{\infty}\frac{2(s+2)}{2n+2+s}
  \left[ \frac{(2n+1)!!}{(2n+2)!!}\right]^{2}
  \left( \frac{w}{w_{0}} \right)^{2n} \ ,
\end{equation}
which claims $\Psi$ to be positive everywhere on the inner equatorial plane 
$w<w_{0}$ between the disk and the black hole. Hence, the critical field 
line of $\Psi=0$ should thread the disk at some point which we denote by 
$w=w_{c}$ (or $r=r_{c}$). For example, if we consider the model of $s=2$ 
with $K$ of the form
\begin{equation}
  K \ = \ 
  \frac{3}{y^{3}}\left( \sqrt{1+y^{2}}-1 \right) 
 -\frac{1}{y}\left( \frac{1}{2}+\frac{1}{\sqrt{1+y^{2}}} \right) \ ,
\end{equation}
the numerical estimation of equation (32) gives $w_{c}\simeq 1.9w_{0}$. 
(The critical field line also reaches to the polar axis at $w\simeq 
1.8w_{0}$, along which $w$ is approximately constant.)

To give the magnetic stream function approximately valid in the whole range 
from $w\sim 1$ to $w\sim w_{0}$, we modify equation (32) which reduces to 
$\Psi\rightarrow \pi B_{0}(Mw)^{2}(1-x^{2})$ in the limit $w\ll w_{0}$. We 
note that $\Psi$ given by equation (32) is smoothly matched to the 
cylindrical field (36) which is valid near the event horizon, only by 
multiplying the factor $(w+1)^{2}/w^{2}$ which becomes equal to unity in 
the region $w\gg 1$. Hence, the modified form of the magnetic stream 
function should be
\begin{equation}
  \Psi \ = \ 
  4(s+2)B_{0}(w+1)^{2}(w_{0}/w)M^{2}\int_{|x|}^{1}dx'
  \left[ \int_{0}^{\pi}d\varphi K(y) \right]
\end{equation}
with $K$ given by equation (35).



We have arrived at the main result (41) available as a model for describing 
the global structure of the black hole magnetosphere. The numerical example 
of magnetic field lines for $s=2$ and $w_{0}=10$ is drawn in Figures~1 
and~2, which are useful to see the magnetic connection between the disk 
and the black hole in the inner region and the conical shape at large $r$, 
respectively. The magnetic spot in the disk connected with the black hole 
is found to be a very narrow region. This is because $\Psi\sim B_{0}M^{2}$ 
near the event horizon, while $\Psi\sim B_{0}(Mw_{0})^{2}$ on the disk 
unless $w\simeq w_{c}$. If the width of the magnetic spot is denoted by 
$\Delta w_{c}$ (or $\Delta r_{c}$), at $w=w_{c}$ we obtain
\begin{equation}
  B_{0}M^{2} \ \sim \ \left( \frac{\partial\Psi}{\partial w} \right)
                      \Delta w_{c} \ ,
\end{equation}
which leads to $\Delta w_{c}\sim 1/w_{0}\ll 1$ (i.e., $\Delta r_{c}\ll M$). 
Because the area $2\pi r_{c}\Delta r_{c}$ of the magnetic spot is just of 
order of the horizon area $16\pi M^{2}$, no significant amplification of 
the magnetic field occurs near the event horizon. Steady inflows of the 
disk plasma to the black hole may occur along the magnetic field lines, 
starting from the magnetic spot of the disk which is apart from the inner 
edge. Then, the inflows reach to the polar region at a distance roughly 
equal to $Mw_{0}$ and finally fall to the black hole along the cylindrical 
field lines.

Plasma outflows from the outer part of the disk may also emanate along the 
open field lines of $\Psi<0$ extending to infinity. It should be remarked, 
however, that the co-existence of two topologically distinct zones in the 
magnetosphere is not due to a general relativistic effect of the black 
hole, because such a global structure is also possible for an appropriate 
superposition of a dipole and a uniform field, giving closed and open field 
lines, respectively. The role of the black hole is rather to prohibit the 
presence of a dipole field at the event horizon. Hence, the closed field 
lines should be sustained by disk currents only, which is the point 
essential to our modeling.

In astrophysical magnetospheres various MHD processes are expected to 
become important for producing highly energetic phenomena. For example, in 
the region very close to the event horizon, strong gravity forces the 
plasma to accrete in radial direction and to bend the cylindrical field 
lines \citep{pu91b,hir92}. The magnetic field (41) generated by disk 
currents provides a starting point to analyze such MHD effects of ambient 
and disk plasma on the magnetospheric structure, which will be discussed in 
the next section.

\section{DISCUSSION}

If the plasma inertia is taken into account even under the condition of 
magnetic domination, the asymptotic collimation from the conical shape of 
field lines should occur in logarithmic scales of the cylindrical radius 
\citep{chi91,tom94}. Apparently such an inertial back-reaction of the 
plasma on the poloidal field does not work in the vacuum fields. Then, we 
present a model with collimated structure, by adding the uniform field 
(which is a solution of the vacuum Maxwell equations)
\begin{equation}
  \Psi_{u} \ = \ \pi B_{u}r^{2}\sin^{2}\theta
\end{equation}
to $\Psi$ given by equation (41). Let us denote the new magnetic stream 
function by $\Psi'(=\Psi+\Psi_{u})$. Because $\Psi$ is negative in the far 
distant region, $B_{u}$ should be chosen to be negative for allowing a 
smooth collimation of the conical field lines. If the uniform field is 
weak, namely $B_{0}+B_{u}>0$, the value of $\Psi'$ remains positive at the 
event horizon to keep the magnetic connection between the disk and the 
black hole (Fig.~3). However, if $B_{0}+B_{u}$ becomes negative, the field 
lines threading the black hole can reach to infinity only (Fig.~4).



Of course, this addition of a uniform field is just a mathematical 
procedure, and the strong collimation shown in Figures~3 and~4 is not a 
consequence of the physical confinement mechanisms by self-generated 
toroidal fields and/or by external pressure with a boundary at finite 
cylindrical radius. It remains quite uncertain whether the details of the 
field-line shape shown in these figures become approximately valid in 
magnetically dominated wind regions or not. Nevertheless, it is interesting 
to note that the vacuum models can successfully describe the transition of 
the structure of field lines threading the black hole (i.e., from magnetic 
connection with the disk to that with the remote load), which may occur in 
more realistic magnetospheres as an observable change of the astrophysical 
activity.

To discuss the magnetic activity by using the vacuum field $\Psi'$ (or 
$\Psi$) as a background field perturbed by magnetically dominated winds 
from the disk, let us consider the magnetosphere slowly rotating with the 
angular velocity $\Omega_{F}$ of a magnetic field line and the hole's one 
$\Omega_{H}$ (i.e., $r\sin\theta\Omega_{F}\ll 1$ and $M\Omega_{H}\simeq 
a/4M\ll 1$). If a magnetic field line threading the black hole is closed in 
the disk at $r=r_{d}$ (see Figs.~1 and~3), the angular velocity 
$\Omega_{F}$ is nearly equal to the local Keplerian one, i.e.,
\begin{equation}
  \Omega_{F} \ \simeq \ \Omega_{K} \ \equiv \ 
                        \left( M/r_{d}^{3}\right)^{1/2} \ ,
\end{equation}
and we have $r_{d}\simeq M\chi w_{0}$, where $\chi$ is a numerical factor 
in the range $1<\chi\leq 1.9$ for the $s=2$ model. Now the approximation of 
slow rotation is assured by virtue of the estimation such that 
$r_{d}\Omega_{F}\simeq1/\sqrt{\beta w_{0}}\ll1$. However, the inner edge in 
realistic accretion disks cannot be too distant from the event horizon 
(namely, $w_{0}$ cannot be too large), and from the relation
\begin{equation}
  \Omega_{H}/\Omega_{F} \ \simeq \ M\Omega_{H}(\beta w_{0})^{3/2} \ ,
\end{equation}
we expect to be $\Omega_{H}<\Omega_{F}$ as a typical case for slowly 
rotating black holes. Hence, the inward winds from the magnetic spot carry 
angular momentum from the disk to the black hole, and the disk plasma which 
remains in the magnetic spot without flowing into the magnetic tube will be 
torqued and fall inward. This means that the disk currents located in the 
inner region and sustaining the loop field lines become unstable for radial 
accretion, and the closed loop field disconnected from the disk must 
disappear owing to the infinite redshift effect at the event horizon 
\citep{pc89}.

In the magnetically-dominated wind region the toroidal magnetic field 
$B_{T}$ is approximately given by
\begin{equation}
  B_{T} \ \simeq -4\pi\eta L/\alpha\varpi \ ,
\end{equation}
where $\alpha$ and $\varpi$ are the lapse function and the cylindrical 
radius, respectively, in Kerr geometry. By virtue of the approximation of 
slow rotation we obtain $\alpha=\sqrt{1-(2M/r)}$ and $\varpi=r\sin\theta$. 
Along a poloidal field line the particle flux $\eta$ per unit flux tube and 
the total angular momentum $L$ of magnetized fluid are conserved, and the 
angular momentum flux per unit area carried by the winds towards the black 
hole is given by $-\eta LB_{p}$, which is a consequence of exerting the 
magnetic torque on the disk plasma. (Now $L$ is positive, while $\eta$ is 
negative for ingoing winds.) Then, the characteristic timescale $t_{1}$ to 
extract the plasma angular momentum in the disk would be
\begin{equation}
  t_{1} \ \sim \ h\rho_{d}r_{d}^{2}\Omega_{K}/(-\eta LB_{p}) \ \sim \ 
  (r_{d}/h)/\beta_{e}\Omega_{K} \ ,
\end{equation}
where $\rho_{d}$ is the mass density in the disk with the thickness $h$, 
and $\beta_e$ is the ratio of the external magnetic stress $B_{T}B_{p}$ at 
the disk surface to the local pressure in the disk. One may also consider 
the internal torque transporting angular momentum radially outward through 
the disk, for which the accretion timescale $t_{2}$ is estimated to be
\begin{equation}
  t_{2} \ \sim \ (r_{d}/h)^{2}/\beta_{i}\Omega_{K} \ ,
\end{equation}
and the ratio $\beta_{i}$ of the internal stress to the local pressure is 
usually supposed to be less than unity (e.g., $\beta_{i}\simeq 0.1$). In 
our modeling the external torque responsible for carrying angular momentum 
into the winds should be dominant if compared with the internal torque, and 
we require the condition $\beta_{e}\gg\beta_{i}h/r_{d}$, which becomes 
valid for $h\ll r_{d}$, unless the turbulent internal disk field is 
supposed to be much stronger than any ordered vertical field at the disk 
surface. Hence, for $\beta_{e}\approx\beta_{i}$, the duration of the closed 
loop structure could be quite longer than the dynamical timescale 
$\Omega_{K}^{-1}$ particularly in a thin disk.

In significant long term accretion of plasma beyond the timescale $t_{1}$ 
one must consider the change of the closed field-line topology, unless 
dynamo action in the inner disk region can efficiently work to recover the 
closed loops. The process of disruption of the closed loops would be 
quasi-stationary, and our models could give a qualitative picture of the 
structure change as follows: Initially strong disk currents can sustain 
nearly circular loops of field lines shown in Figure~1 (i.e., 
$B_{0}\gg|B_{u}|$). As the disk currents become weaker (i.e., as the ratio 
$B_{0}/|B_{u}|$ decreases) by virtue of the radial accretion, the loops 
begin to collapse towards the black hole (especially in vertical 
direction), as was shown in Figure~3. Finally, only the loops disconnected 
from the horizon remains near the inner edge (see Fig.~4 for 
$B_{0}\leq|B_{u}|$). The loop field completely annihilates as a set of 
O-points, if the radial accretion from the inner edge continues still more 
\citep{pc90b}. During this process, a part of electromagnetic energy stored 
in loop field lines could be outward transferred as a Poynting flux to 
infinity. We expect that the magnetic field (41) to become a convenient 
initial state for analytical and numerical MHD calculations to study the 
problem of non-steady energy release in plasma accretion toward the black 
hole (see, e.g., \citet{ko00} for numerical investigations, in which the 
initial magnetic field is chosen to be the Wald solution in Kerr 
background).

In the slowly rotating magnetospheres the closed field line plays the role 
of a path for the energy and angular momentum transfer from the disk to the 
black hole. [ \citet{li00c} claimed that the transfer from the black hole to 
the disk occurs only for $a/M>0.36$.]  However, the spin-up of the black 
hole should stop when the structure shown in Figure~4 has been established. 
Then, the energy extraction from the rotating black hole becomes possible 
as a mechanism of energy release. The angular velocity $\Omega_{F}$ of the 
open field lines connecting the black hole with the remote load may satisfy 
the condition $\Omega_{F}\approx\Omega_{H}/2$, which allows the twisted 
field with the negative toroidal component $B_{T}$ (i.e., $\eta<0$ and 
$L<0$) to bring negative angular momentum and energy to the black hole. The 
efficiency of this Blandford-Znajek mechanism is still controversial, 
because the validity of the condition $\Omega_{F}\approx\Omega_{H}/2$ is 
based on the assumption that a causal contact between the black hole and 
the remote plasma holds. In plasma accretion onto the black hole the 
poloidal inward velocity exceeds the fast magnetosonic speed near the 
horizon, and no information could propagate outward across some critical 
surface \citep{pc90a}. In fact, the black hole-driven winds satisfying 
$\Omega_{F}\ll\Omega_{H}$ have been found to exist \citep{pu98}. For such 
winds the main mechanism of energy extraction from the black hole could be 
the screw instability, which makes the twisted magnetic tubes unwind and 
causes a sudden transport of the magnetic free-energy into the plasma's 
kinetic energy. For the cylindrical magnetic field shown in Figure~4 we can 
use the instability condition known as the Kruskal-Shafranov criterion 
\citep{kad66} of the form
\begin{equation}
  |B_{T}|/2\pi\delta \ > \ |B_{p}|/S \ ,
\end{equation}
where $\delta$ and $S$ are the width and the length of the cylindrical 
magnetic tube. This has been proved by \citet{gru99} in the force-free 
magnetosphere, but it should be noted that the analysis has been limited to 
the case such that $\Omega_{F}=0$ and $S\gg M$ (i.e., the approximation of 
no rotation and no gravity). Then, a naive application \citep{li00b} of the 
Kruskal-Shafranov criterion to the magnetosphere where $B_{T}$ is roughly 
estimated to be $r\sin\theta\Omega_{F}B_{p}$ seems to be very dubious. 
Here, the toroidal field is estimated by equation (46), which remains valid 
even for $\Omega_{F}=0$, and considering the region well away from the 
event horizon (i.e., $\alpha\simeq1$), we rewrite the instability condition 
into the form
\begin{equation}
  S \ > \ |B_{p}|\delta^{2}/(2|\eta L|) \ ,
\end{equation}
of which the right-hand side is constant if the magnetic flux 
$|B_{p}|\delta^{2}$ in the cylinder is conserved. This means that if a 
larger angular momentum flux $|\eta LB_{p}|\delta^{2}$ is carried away from 
the black hole by magnetically dominated winds in the cylinder with the 
width $\delta\sim M$, the twisted field lines become screw-unstable in a 
shorter scale $S$. After the disruption of the screw-unstable structure of 
magnetic field, the rotating black hole would begin to twist again the 
magnetic field lines, and the energy release from the black hole could 
quasi-periodically occur.

As was first proposed by \citet{gru99}, the closed field lines connecting 
the black hole with the disk may be also screw-unstable. However, for the 
field lines closed in the disk the toroidal component will be mainly 
generated by disk rotation with the angular velocity 
$\Omega_{K}\simeq\Omega_{F}$. Then, we must discuss the instability 
condition by including the effect of $\Omega_{F}$, which is a problem 
studied in future works \citep{ma01}.

Finally it should be emphasized that the closed field-line topology 
presented in this paper is a very transient structure in rapidly rotating 
plasma-filled magnetospheres. The angular velocity 
$\Omega_{F}\simeq\Omega_{K}$ of closed loops connecting the black hole with 
the disk will be smaller than the hole's angular velocity $\Omega_{H}$, and 
the energy and angular momentum transfer from the black hole to the disk 
could spin up the disk plasma \citep{li00c}. Then, as a consequence of this 
energy extraction from the black hole, outward flows of plasma from the 
disk would be generated, and owing to the frozen-in condition the field 
lines threading the black hole are eventually disconnected from the disk. 
The closed field lines anchored in plasma orbiting in the ergosphere would 
also open up to infinity by the strong dynamo action, as was conjectured to 
occur in field-aligned pulsars \citep{pc90b}. Further, the timescale 
$t_{2}$ for radial accretion due to internal torque may become shorter than 
the timescales to open up the loops closed in the disk. Then, the closed 
loops accrete toward the black hole and annihilate owing to the infinite 
redshift effect \citep{pc90b}. These are astrophysically interesting 
processes of rapid energy release, and we could observe transient flaring 
states of black hole magnetospheres, if the repeated formation of closed 
magnetic loops due to strong disk currents in the inner region occurs.

\acknowledgments

We are grateful to T. Matsuoka for valuable discussion and to an anonymous 
referee for strong criticism which helped us to improve the paper .

\clearpage

%
%
%


\begin{figure}
    \epsscale{0.9}
 \begin{center}
    \plotone{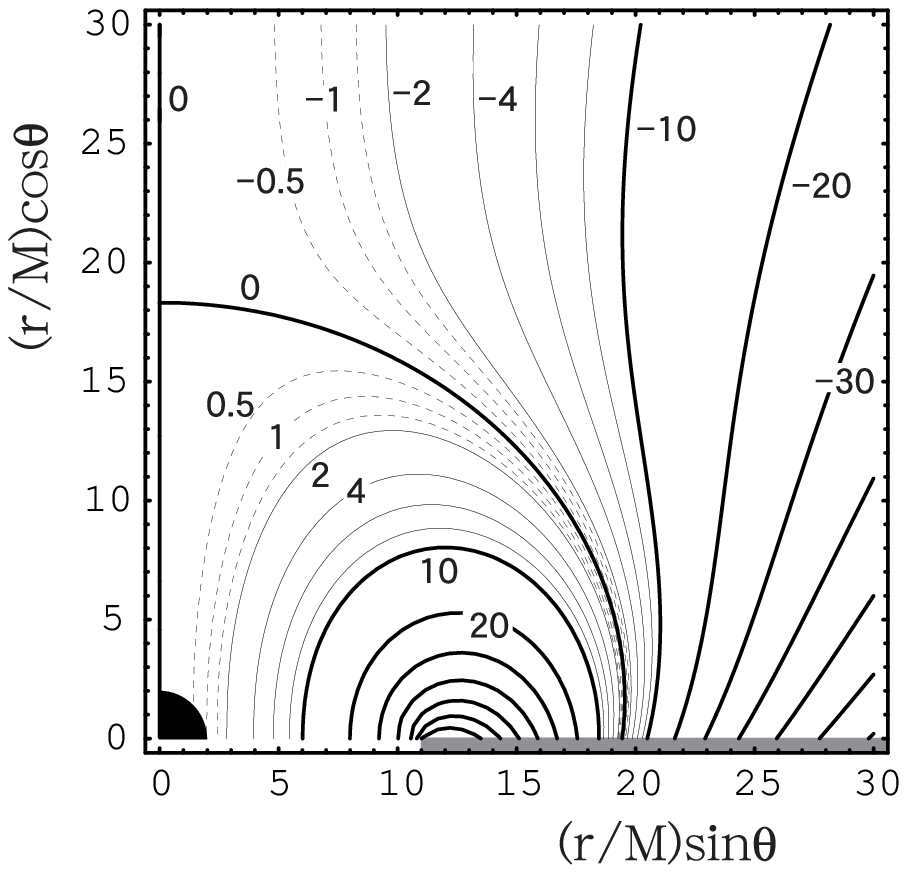}
    \caption{
Poloidal magnetic field lines for the axisymmetric 
model of $s=2$ and $w_{0}=10$. The value of $\Psi/4\pi M^{2}B_{0}$ is 
denoted on each field line, and the structure is assumed to have the 
equatorial symmetry. Further, the central black hole with the radius $r=2M$ 
and the surrounding disk with the inner edge at $r=11M$ are displayed. We 
can see the magnetic connection of the black hole with a very narrow region 
in the disk.
 }
 \end{center}
\end{figure} 

\begin{figure}
    \epsscale{0.9}
 \begin{center}
    \plotone{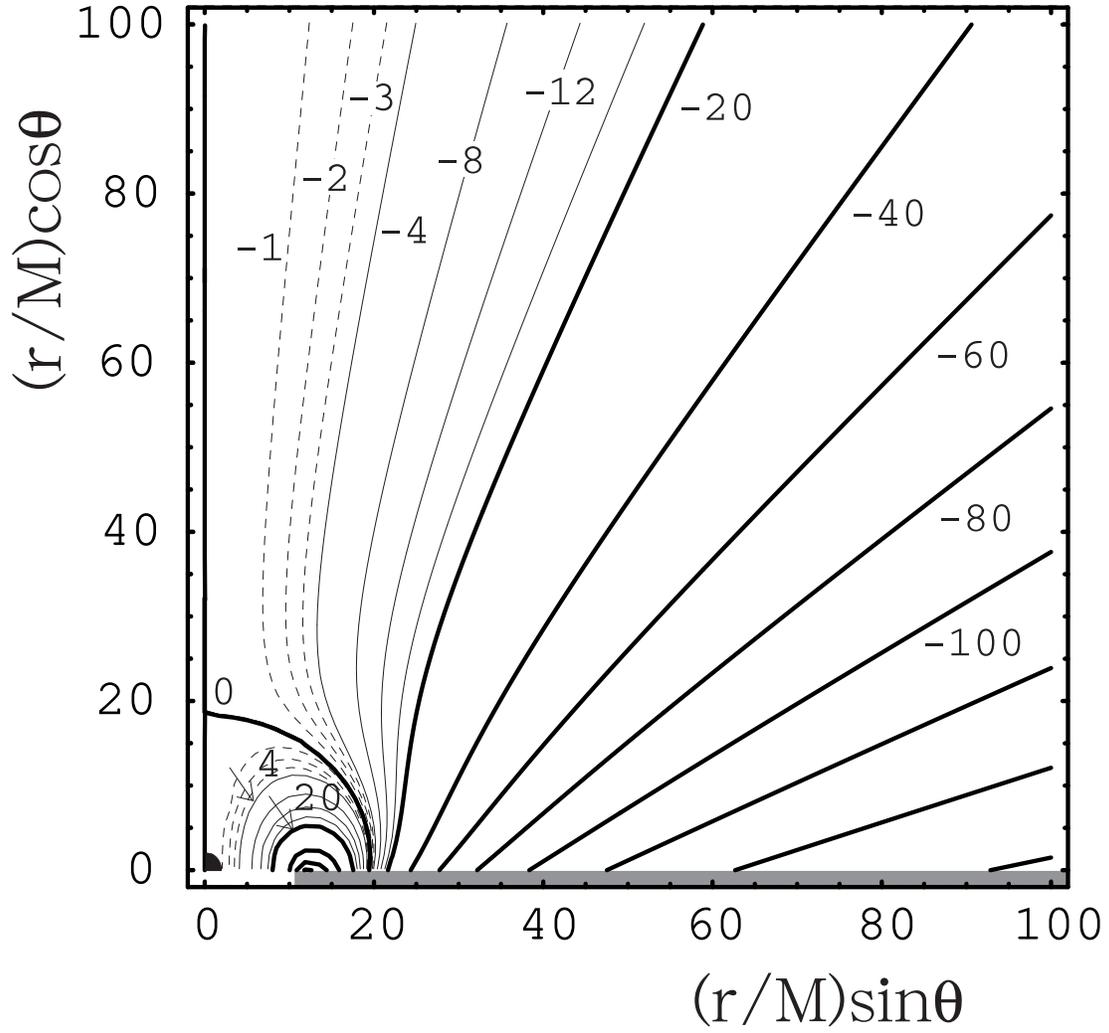}
    \caption{
Same as Fig.~1 for large scale of $r$, to show the 
asymptotic shape of field lines.
 }
 \end{center}
\end{figure} 

\begin{figure}
    \epsscale{0.9}
 \begin{center}
    \plotone{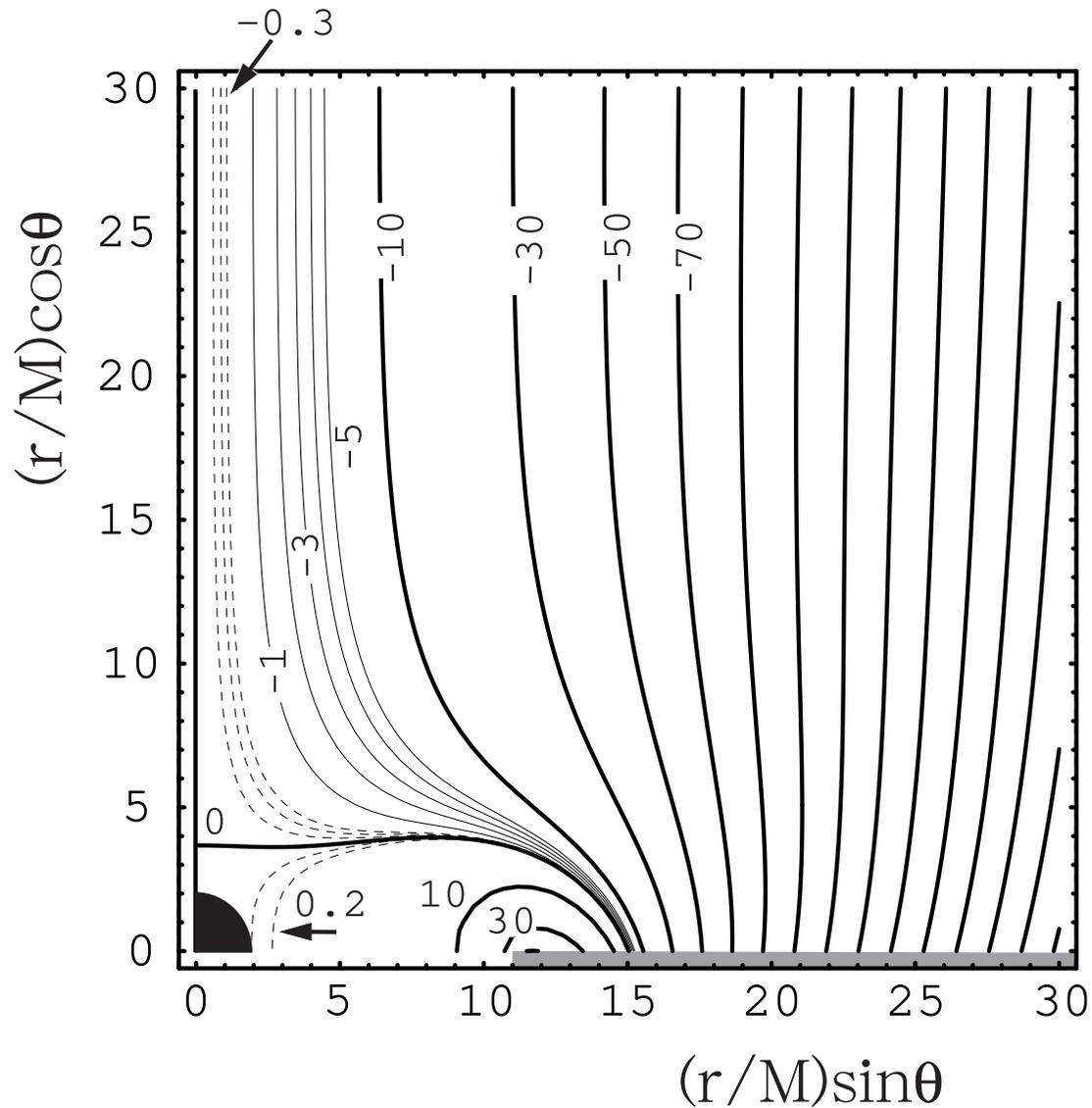}
    \caption{
Poloidal magnetic field lines for the model $\Psi'$ 
involving a weak uniform field, i.e., $B_{u}=-0.9B_{0}$. The magnetic field 
generated by the disk current is same as Fig.~1. The magnetic connection 
between the black hole and the disk still remains, and the collimated field 
lines appear in large $r$.
 }
 \end{center}
\end{figure} 

\begin{figure}
    \epsscale{0.9}
 \begin{center}
    \plotone{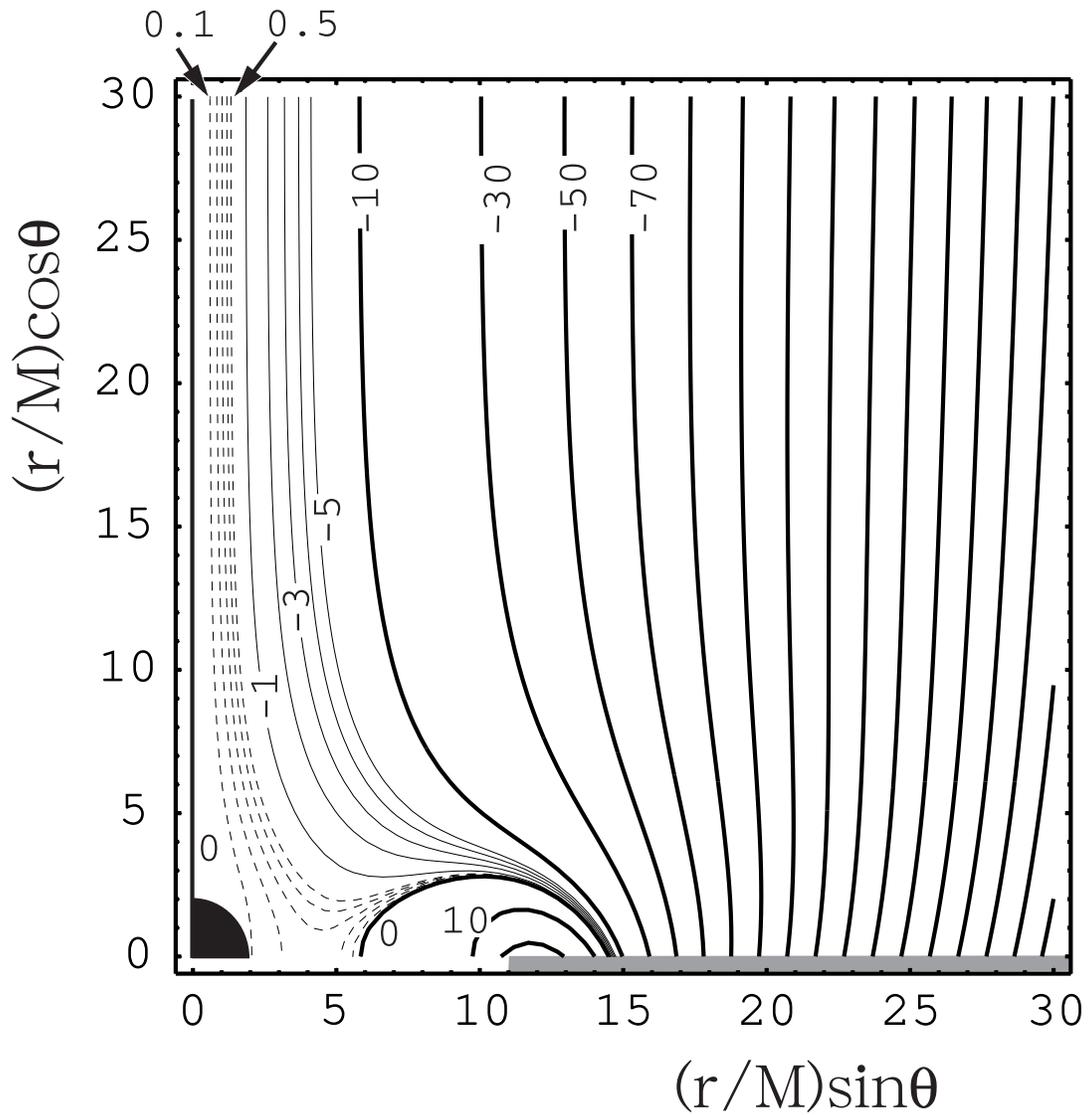}
    \caption{
Same as Fig.~3, except for the existence of a strong 
uniform field, i.e., $B_{u}=-1.1B_{0}$. Only the open field lines can 
thread the black hole without any connection with the disk, though the loop 
field lines still exist near the inner edge. 
 }
 \end{center}
\end{figure} 

\end{document}